\documentstyle[aps,multicol,epsfig]{revtex}

\renewcommand{\narrowtext}{\begin{multicols}{2} \global\columnwidth20.5pc}
\renewcommand{\widetext}{\end{multicols} \global\columnwidth42.5pc}

\def\top#1{\vskip #1\begin{picture}(290,80)(80,500)\thinlines \put(65,500){\line( 1, 0){255}}\put(320,500){\line( 0, 1){5}}\end{picture}}
\def\bottom#1{\vskip #1\begin{picture}(290,80)(80,500)\thinlines \put(330,500){\line( 1, 0){255}}\put(330,500){\line( 0, -1){5}}\end{picture}}

\newcommand{\bq}{\begin{equation}}
\newcommand{\eq}{\end{equation}}
\newcommand{\bqa}{\begin{eqnarray}}
\newcommand{\eqa}{\end{eqnarray}}
\newcommand{\nn}{\nonumber \\}

\begin{document}
\draft 
\title{Doping and temperature dependence of superfluid weights for high $T_c$ cuprates} 

\author{Sung-Sik Lee and Sung-Ho Suck Salk}
\address{Department of Physics, Pohang University of Science and Technology,\\
Pohang, Kyoungbuk, Korea 790-784\\}
\date{\today}

\maketitle

\begin{abstract}
Using the improved U(1) and SU(2) slave-boson approaches of the t-J Hamiltonian [Phys. Rev. B {\bf 64}, 052501 (2001)] that we developed recently, we report the doping and temperature dependence of the superfluid weight. 
It is shown that at low hole doping concentration $x$ and at low temperatures $T$ there exists a propensity of a linear decrease of the superfluid weight $n_s/m^*$ with temperature, and a tendency of doping independence in the linearly decreasing slope of $\frac{n_s}{m^*}(x,T)$ vs. $T$ in agreement with the experimentally observed relation $\frac{n_s}{m^*}(x,T) = \frac{n_s}{m^*}(x,0) - \alpha T$ with $\alpha$, a constant.
It is also demonstrated that both $T_c$ and $n_s/m^*$ increase with hole doping concentration $x$ in the underdoped region, reaches a saturation(maximum) at a hole doping above optimal doping and decreases beyond the saturation point in the overdoped region.
Such a reflex (decreasing) behavior of $T_c$ and $n_s/m^*$ is attributed to the weakening of coupling between the spin(spinon pair order) and charge(holon pair order) degrees of freedom in the overdoped region.
All of these findings are in agreement with $\mu$SR measurements.
\end{abstract}
\narrowtext

\newpage
Presently there exist two outstanding physical problems of superfluid weights to be resolved.
One is the understanding of the so-called `boomerang' (reflex) behavior and the other, the doping independence of linear decrease of superfluid weight $\frac{n_s}{m^*}(x,T)$ vs. temperature $T$.
More than a decade ago the transverse field muon-spin-relaxation($\mu$-SR) measurements of the magnetic penetration depth $\lambda$ in high $T_c$ copper oxide superconductors revealed the `boomerang' behavior 
of the superconducting temperature $T_c$ and the superfluid weight $n_s/m^*$, 
that is, a linear increase of both $T_c$ and $n_s/m^*$ in the underdoped region, a saturation of $T_c$ at optimal doping and a decrease (reflex) of both $T_c$ and $n_s/m^*$ in the overdoped region as the hole doping concentration increases\cite{UEMURA,UEMURA_NATURE,BERNHARD}.
However, theoretical explanations of this boomerang (reflex) behavior are still incomplete.
Most recently, magnetic penetration depth($\lambda$) measurements\cite{SHENG,PANA0,PANA} showed a doping independence in the linearly decreasing slope of the superfluid weight with increasing temperature obeying the relations $\frac{n_s}{m^*}(x,T) = \frac{n_s}{m^*}(x,0) - \alpha T$ in the underdoped region where $x$ is the hole concentration and $\alpha$, a constant.
In these experiments\cite{SHENG,PANA0,PANA}, the normalized superfluid weight shows a universal behavior as a function of scaled temperature, that is, $\frac{n_s/m^*(x,T)}{n_s/m^*(x,0)} = 1 - \beta (T/T_c)$.
This universal behavior has been interpreted by the phase fluctuation model\cite{CARLSON}, or the quasiparticle model based on the Landau-Fermi liquid theory\cite{MILLIS,MESOT,BENFATTO}.
However, apart from these phenomenological models the theoretical explanation of this universal behavior from a microscopic theory is largely unaccomplished\cite{ORENSTEIN}.

There have been numerous efforts to theoretically explain the microscopic origin for this doping independent propensity of the decreasing slope of the observed superfluid weight with increasing temperature\cite{PALEE97,DHLEE,WANG}.
However, until now the problem still remains unexplained\cite{ORENSTEIN}.
Earlier U(1) slave-boson theories\cite{PALEE97,DHLEE,KOTLIAR,FUKUYAMA,UBBENS} of the $t-J$ Hamiltonian predicted that at small hole doping the superfluid weight $\frac{n_s}{m^*}(x,T)$ as a function of temperature is strongly doping ($x^2$) dependent.
This is in direct conflict with the $\mu$SR measurements\cite{SHENG,PANA0,PANA}.
Lee and Wen\cite{PALEE97,WEN} proposed that the SU(2) theory which incorporates the low energy phase fluctuations of order parameters may resolve this problem.
Recently D.-H. Lee\cite{DHLEE} presented a U(1) slave-boson theory concerned with the low energy fluctuations of order parameters and the excitations of massless Dirac fermions at the d-wave nodal point.
This theory, also, showed the $x^2$ dependence of $\frac{n_s}{m^*}$ at finite temperature. 
Most recently Wang et al.\cite{WANG} also showed that the recently proposed d-density wave theory\cite{NAYAK} fails to predict the doping independence of the linearly decreasing slope of superfluid density vs. temperature.

Earlier, we presented an improved U(1) and SU(2) slave-boson approach of the t-J Hamiltonian which differs from other slave-boson approaches to the t-J Hamiltonian in that coupling between the charge and spin degrees of freedom is manifested in the slave-boson representation of the Heisenberg term of $J\sum_{<i,j>} ( {\bf S}_{i} \cdot {\bf S}_{j} - \frac{1}{4}n_{i}n_{j} )$, by naturally admitting the charge degree of freedom exhibited in the intersite charge coupling energy term of $\frac{1}{4}n_{i}n_{j}$\cite{LEE}.
In this theory\cite{LEE} holons (charges) are subject to attractive interaction as a result of coupling to the spinon singlet pairs.
As a result, holons form holon pairs and the holon-pair bosons undergo bose condensation\cite{NOZIERE} at low temperatures.
This is the major difference of the present holon-pair boson theory as compared to other earlier theories which pay attention to singe-holon bose condensation\cite{PALEE97,DHLEE,KOTLIAR,FUKUYAMA,UBBENS,WEN}.
Our theory was shown to predict the arch shaped superconducting (holon-pair bose condensation) temperature $T_c$ as a function of doped hole concentration $x$\cite{LEE}, in agreement with the experimentally observed phase diagram.
It is, thus, of great interest to employ this theory for checking its predictability of the boomerang and both the doping and temperature dependence of the superfluid weight.

Since the predicted results of various physical properties are qualitatively indistinguishable between the U(1) and SU(2) slave-boson theories, below we present only the SU(2) theory for generality.
In the SU(2) slave-boson representation\cite{WEN}, the electron operator is given by $c_{\alpha}  = \frac{1}{\sqrt{2}} h^\dagger \psi_{\alpha}$ with $\alpha=1,2$, where $\psi_1=\left( \begin{array}{c} f_1 \\ f_2^\dagger \end{array} \right)$,   $\psi_2 = \left( \begin{array}{c} f_2 \\ -f_1^\dagger \end{array} \right)$ and $h = \left( \begin{array}{c}  b_{1} \\ b_{2} \end{array} \right)$ are respectively the doublets of spinon and holon annihilation operators in the SU(2) theory.
The SU(2) slave-boson representation of the t-J Hamiltonian shows\cite{LEE}
\bqa
H  & =  &  - \frac{t}{2} \sum_{<i,j>}  \Bigl[ (f_{\alpha i}^{\dagger}f_{\alpha j})(b_{1j}^{\dagger}b_{1i}-b_{2i}^{\dagger}b_{2j})  + h.c. \nn
&& + (f_{2i}f_{1j}-f_{1i}f_{2j}) (b_{1j}^{\dagger}b_{2i} + b_{1i}^{\dagger}b_{2j}) + h.c. \Bigr] \nn
 && -  \frac{J}{2} \sum_{<i,j>} ( 1 + h_{i}^\dagger h_{i} ) ( 1 + h_{j}^\dagger h_{j} ) \times \nn
 && (f_{2i}^{\dagger}f_{1j}^{\dagger}-f_{1i}^ {\dagger}f_{2j}^{\dagger})(f_{1j}f_{2i}-f_{2j} f_{1i}) -  \mu \sum_i  h_i^\dagger h_i  \nn
  && -  \sum_i  \Bigl[ i\lambda_{i}^{(1)} ( f_{1i}^{\dagger}f_{2i}^{\dagger} + b_{1i}^{\dagger}b_{2i}) + i \lambda_{i}^{(2)} ( f_{2i}f_{1i} + b_{2i}^\dagger b_{1i} ) \nn
  && + i \lambda_{i}^{(3)} ( f_{1i}^{\dagger}f_{1i} -  f_{2i} f_{2i}^{\dagger} + b_{1i}^{\dagger}b_{1i} - b_{2i}^{\dagger}b_{2i} ) \Bigr],
\label{eq:su2_sb_representation}
\eqa
where $\lambda_{i}^{(1),(2),(3)}$ are the real Lagrangian multipliers to enforce the local single occupancy constraint in the SU(2) slave-boson representation\cite{WEN}.
It is noted that the inclusion of the charge-charge interaction $-\frac{J}{4} n_i n_j$ in the Heisenberg Hamiltonian results in the coupling between the spin (spinon) and charge (holon) degrees of freedom.
It is easy to show that under inverse SU(2) transformation of the above expression we recover the original form of the U(1) slave-boson representation given in the literature\cite{LEE}.

After relevant Hubbard-Stratonovich transformations\cite{LEE}, the holon action is obtained to be 
\bqa
&& S^b({\bf A},\chi, \Delta^f, \Delta^B, h) = \int_0^\beta d\tau \Big[
\sum_i h^\dagger({\bf r}_i,\tau) ( \partial_\tau - \mu ) h({\bf r}_i,\tau) \nn
&& + \frac{J}{2} \sum_{<i,j>} |\Delta^f_{ij}|^2 \Big( \sum_{\alpha,\beta} |\Delta^b_{ij;\alpha \beta}|^2 + x^2 \Big) \nn
&& -\frac{t}{2} \sum_{<i,j>} \Big( e^{i A_{ij}} h^\dagger({\bf r}_i, \tau) U^b_{ij} h({\bf r}_j, \tau) + c.c. \Big) \nn
&& -  \frac{J}{2} \sum_{<i,j>} \Big( |\Delta^f_{ij}|^2 h^\dagger({\bf r}_i, \tau) \Delta^B_{ij} (h^\dagger({\bf r}_j, \tau))^T + c.c. \Big)
\Big],
\label{eq:su2_holon_action}
\eqa
and the spinon action,
\bqa
&& S^f(\chi, \Delta^f, \psi) = \int_0^\beta d\tau \Big[
\sum_i \psi^\dagger({\bf r}_i, \tau)  \partial_\tau  \psi({\bf r}_i, \tau) \nn
&&  + \frac{J(1- x^2)}{2} \sum_{<i,j>} \Big( |\Delta^f_{ij}|^2 + \frac{1}{2}|\chi_{ij}|^2 \Big) \nn
&& -\frac{J}{4}(1- x )^2 \sum_{<i,j>} \Big( \psi^\dagger({\bf r}_i, \tau) U^f_{ij} \psi({\bf r}_j, \tau) + c.c. \Big)
\Big].
\label{eq:su2_spinon_action}
\eqa
Here various symbol definitions are as follows.
$h({\bf r}_i,\tau) = \left( \begin{array}{c} b_{1}({\bf r}_i,\tau) \\ b_{2}({\bf r}_i,\tau) \end{array} \right)$ is the SU(2) doublet of holon field, and $\psi({\bf r}_i, \tau) = \left( \begin{array}{c} f_{1}({\bf r}_i,\tau) \\ f_{2}^\dagger({\bf r}_i,\tau) \end{array} \right)$, the SU(2) doublet of spinon field.
${\bf A}$ is the electromagnetic(EM) field.
$ U^b_{i,j}  =  \left( \begin{array}{cc} \chi^*_{ij} & - \Delta^f_{ij} \\
                           - \Delta^{f*}_{ij} & -\chi_{ij}
                         \end{array} \right)$ and
$ U^f_{i,j}  =  \left( \begin{array}{cc} \chi^*_{ij} & - 2\Delta^f_{ij} \\
                           - 2\Delta^{f*}_{ij} & -\chi_{ij}
                         \end{array} \right)$ 
are respectively the order parameter matrices of hopping($\chi_{ij}$) and spinon pairing($\Delta^f_{ij}$). 
Here $\chi_{i,i+l} = \eta e^{i \alpha_{l}({\bf r}_i) } \cos (\theta_{l}^0 + \theta_{l}({\bf r}_i)) $ and $\Delta^f_{i,i+l} = \eta e^{i \beta_l({\bf r}_i) } \sin (\theta_{l}^0 + \theta_{l}({\bf r}_i))$, with  $\eta = \sqrt{ | \chi |^2 + | \Delta^f |^2 }$ and $\theta^0_l = \pm \tan^{-1} \frac{\Delta_f}{\chi}$(the sign $+(-)$ is for the ${\bf ij}$ link parallel to $\hat x$ ($\hat y$)).
$\alpha_l({\bf r}_i)$ represents the phase fluctuations of the hopping order parameter, $\beta_l({\bf r}_i)$, the phase fluctuations of the spinon pairing order parameter, and $\theta_l({\bf r}_i)$, the relative phase fluctuations between the amplitudes of hopping and spinon pairing order parameters.
$ \Delta^B_{ij}  =  \left( \begin{array}{cc} \Delta^b_{ij;11} &  \Delta^b_{ij;12} \\
                            \Delta^{b}_{ij;21} & \Delta^b_{ij;22}
                         \end{array} \right)$
is the matrix of the holon pairing order parameter.

After integration over the holon and spinon fields, we obtain the total free energy,
\bqa
F({\bf A}) = -\frac{1}{\beta} \ln \int D\chi D\Delta^f D\Delta^B \nn
e^{-\beta (F^b({\bf A}, \chi, \Delta^f, \Delta^B) + F^f(\chi, \Delta^f)) }, 
\label{eq:free_energy_su2_2}
\eqa
where $F^b({\bf A}, \chi, \Delta^f, \Delta^B ) = -\frac{1}{\beta}\ln \int Dh e^{-S^b({\bf A},\chi, \Delta^f, \Delta^B, h)}$ is the holon free energy and $F^f(\chi, \Delta^f) = -\frac{1}{\beta}\ln \int D\psi e^{-S^f(\chi, \Delta^f, \psi)}$, the spinon free energy.

To compute the EM current response function, we first obtain the total free energy by integrating out the three phase fields $\alpha$, $\beta$ and $\theta$ above.
We obtain a formula for the EM current response function, up to the second order, 
\widetext
\top{-2.8cm}
\bqa
\Pi_{lm}(\omega, {\bf q}) & = & \Pi_{lm}^{b(A, A)}(\omega, {\bf q}) \nn
& - & \sum_{a^1,a^2=\theta, \alpha, \beta} \sum_{l^{'},m^{'}=\hat x,\hat y} \Pi_{ll^{'}}^{b(A,a^1)}(\omega, {\bf q}) \left[ \Pi^{b}(\omega, {\bf q}) + \Pi^{f}(\omega, {\bf q}) \right]^{-1}_{a^1,l^{'};a^2,m^{'}}  \Pi_{m^{'}m}^{b(a^2,A)}(\omega, {\bf q}),
\label{eq:ioffe_larkin_su2}
\eqa
\bottom{-2.8cm}
\narrowtext
\noindent
where $\left[ \Pi^{b}(\omega, {\bf q}) + \Pi^{f}(\omega, {\bf q}) \right]^{-1}_{a^1,l^{'};a^2,m^{'}}$ is the inverse matrix element of $\Big[ \Pi^{b(a^1,a^2)}_{l^{'}m^{'}}(\omega, {\bf q})$ $+$ $ \Pi^{f(a^1,a^2)}_{l^{'}m^{'}}(\omega, {\bf q}) \Big]$.
Here $\Pi_{lm}^{b(a^1,a^2)}$ and $\Pi_{lm}^{f(a^1,a^2)}$ represent the isospin current response functions of holons and spinons respectively to gauge fields $a^1, a^2 = \theta, \alpha,$ or $\beta$.
Similarly, $\Pi_{lm}^{b(A, A)}(\omega, {\bf q})$ is the EM current response function of holon to the EM field.
$\Pi_{lm}^{b(a^1,A)}$ is the isospin current response function of holon to both gauge fields $a$ and $A$.
It is noted that the SU(2) isospin current of holon(spinon) is defined as $j_{\alpha,l}^{b(f)} = -\beta \frac{\delta F^{b(f)}}{\delta a^\alpha_{i,i+l}}$,  and analogously for the `EM' current, $j_l = -\beta \frac{\delta F}{\delta A_l}$.
In the SU(2) slave-boson theory, the response function $\Pi_{ll^{'}}^{b(A,a^1)}$ of the holon isospin current to the EM field vanishes owing to the contribution of the $b_2$-boson in the static and long-wavelength limit\cite{WEN}.
Therefore the superfluid weight of the total system is given only by the holon current response function,
\bqa
\frac{n_s}{m^*} = - \frac{1}{e^2} \lim_{ {\bf q} \rightarrow 0} \Pi_{ll}^{b(A, A)}(\omega=0, {\bf q} ).
\label{eq:su2_superfluie_weight}
\eqa
Here $\Pi_{lm}^{b(A, A)}(\omega, {\bf q})$ is computed from the use of the usual linear response theory for the holon action Eq.(\ref{eq:su2_holon_action})\cite{SCHRIEFFER}.

In Fig. 1(a), with the use of our slave-boson theory\cite{LEE} we show the temperature dependence of superfluid weight for a wide range of doping concentrations covering both underdoping and overdoping.
The slope of each curve represents the variation of superfluid weight with temperature which is not in precise agreement with observations\cite{SHENG,PANA0,PANA}, overestimating the superfluid weight $\frac{n_s}{m^*}(x,0)$ at $T = 0 K$ and further $\frac{n_s}{m^*} (x,T)$ reaches zero rather abruptly compared to the observation which showed a slower drop\cite{PANA0}.
We would like to point out that a close look at Fig. 1 of the above experimental paper\cite{PANA0} reveals a linear slope behavior only in the lower temperature region and a quick non-linear drop of superfluid weight at the higher temperature region.
Such a change of slope from linearity is qualitatively in agreement with our results although our theory predicted a faster drop in $\frac{n_s}{m^*}(T)$.
The continuous drop in the predicted superfluid weight indicates a second order phase transition.
We believe that improvement can be achieved if we can introduce the $t^{'}$ term in our t-J Hamiltonian which is known to cause a good agreement with the observed dispersion energy by showing realistic cold and hot spots.
The contribution of the hot spot in the Brillouin zone becomes increasingly important with increasing temperature up to $T_c$ to allow classical phase fluctuations.

As temperature increases, the predicted superfluid weight shows a tendency of linear decrease in $T$ with nearly identical slopes particularly at low temperatures in the underdoped region.
To rigorously check such a propensity of doping independence in the slope of $\frac{n_s}{m^*}$ vs. $T$, we computed differences in the superfluid weight at $T=0$ and $T \neq 0$, that is, $\frac{n_s}{m^*}(x,T) - \frac{n_s}{m^*}(x,0)$ at each $x$.
The results are shown in Fig. 1 (b).
A decrease of the superfluid weight with temperature at all dopings is predicted although the temperature dependence of the superfluid weight does not show a clear linearity at all temperatures as shown in Fig. 1 (b).
There exists a tendency of linear decrease with a nearly identical slope particularly in the underdoped region only at low temperature.
Although the temperature dependence of $\frac{n_s}{m^*}(x,T)$ does not show a global linearity by yielding a good fit to the empirical relation of $\frac{n_s}{m^*}(x,T) = \frac{n_s}{m^*}(x,0) - \alpha(x) T - \alpha^{'}(x) T^2 - O(T^3)$, the linear slope $\alpha(x)$ is found to be doping in dependent only in the underdoped region at low temperature as is clearly shown in Fig. 1 (b).
As shown in the figure it is found that the variation of the slope, that is, the superfluid weight $n_s/m^*$ vs. $T/t$ at low temperatures does not appreciably change in the underdoped region ($x=0.04$, $0.07$ and $0.1$).
On the other hand, in the overdoped region ($x=0.16$, $0.18$ and $0.2$) the slope of the superfluid weight with temperature is no longer doping-independent quite unlike the underdoped case, again, in agreement with observation.
The predicted optimal doping is $x_o = 0.13$.
It is noted that there exists a crossing behavior between the underdoped and overdoped lines of superfluid weights vs. temperature as shown in Fig. 1(a).
This prediction is consistent with measurements\cite{SHENG,PANA0,PANA}.

In Fig. 1(c) we display the the normalized superfluid weight ($\frac{n_s/m^*(x,T)}{n_s/m^*(x,0)}$) as a function of the scaled temperature ($T/T_c$) for various hole concentrations.
As shown in the figure, the normalized superfluid weight begins to decrease linearly at low temperature and drops more rapidly at higher temperature, by showing a convex line shape in qualitative agreement with experiments\cite{SHENG,PANA0,PANA}.
The decrease of superfluid weight with temperature occurs as a consequence of diminishing spin singlet pairing order with increasing temperature.
This is readily understood from the holon pairing term (the last term in Eq. (2)) which reveals the interplay of the spin (spinon) singlet pairing order with the holon pairing order;
the spin singlet pairing order is predicted to decrease with temperature.
Thus the decrease of the superfluid weight with increasing temperature occurs by breaking the holon pairs (hole pairs) in connection with the diminishing singlet pairing order at higher temperatures.
A linear decrease of superfluid weight is predicted at low temperature and a rapid decrease at higher temperature, displaying a convex line shape. 
This trend of the convex line shape for all doping cases is in complete agreement with observation\cite{SHENG,PANA0,PANA}.
Although the predicted $\beta$ value in the relation of $\frac{n_s/m^*(x,T)}{n_s/m^*(x,0)} = 1 - \beta (T/T_c)$ shows linear decrease particularly in the lower temperature region in agreement with observation, its $\beta$ value scattered around $0.055$ fails to quantitatively agree with the empirical value of $\beta \approx 0.5$ independent of doping in the underdoped region \cite{PANA}. 
Such discrepancy indicates that the inclusion of the next nearest hopping ($t^{'}$) term is important.
This is because the Fermi surface segment grows from the nodal point initially at half-filling and closes near $(\pi, 0)$ as hole doping reaches optimal doping and beyond as is well-known from the angle resolved photoemission spectroscopy measurements\cite{MARSHALL}.
In the future the inclusion of the $t^{'}$ term is necessary to take a good account of classical phase fluctuations in the region of the hot spot at high temperatures near and at $T_c$

According to the present theory the superconductivity is accomplished by the condensation of the Cooper pairs of d-wave symmetry as composites of the holon pairs of s-wave symmetry and the spinon pairs of d-wave symmetry.
Thus the Cooper pairs can be regarded as composite objects resulting from such coupling of the spin (spinon pairing) and charge (holon-pairing) degrees of freedom.

To clearly show the effect of coupling between the spin(spinon) and charge(holon) degrees of freedom, we calculated the superfluid weight by varying the antiferromagnetic spin-spin coupling, that is, the Heisenberg coupling $J$.
In Fig.2, we show the $J$ dependence of superfluid weight at $x=0.1$ for three different choices of $J/t=0.2, 0.3$ and $0.4$. 
As $J$ increases, a decreasing tendency in the slope of the superfluid weight vs. temperature is predicted.
For stronger Heisenberg coupling $J$ and thus the spinon pairing strength (order), it is more difficult to thermally break the holon-pairs due to a larger holon-spinon coupling.
Such coupling effect is manifested in the last term of Eq.(\ref{eq:su2_holon_action}).
This results in the more slowly decreasing slope of the superfluid weight with increasing temperature for large $J$, as shown in the figure.

In Fig.3, we show the predicted boomerang behavior in the plane of $T_c$ vs. $\frac{n_s}{m^*}(x,T \rightarrow 0)$, by showing a linear relationship between the two in the underdoped region and the `reflex' behavior in the overdoped region based on the U(1) theory.
By the reflex behavior we mean the decrease of both $T_c$ and $\frac{n_s}{m^*}$ as the hole doping concentration increases in the overdoped region.
This predicted trend is consistent with the measurements\cite{UEMURA,UEMURA_NATURE,BERNHARD} of the muon-spin-relaxation rates $\sigma$ ($\sigma \propto n_s/m^* \propto T_c$).
The pseudogap temperature is predicted to rapidly decrease in the overdoped region in agreement with observation.
The predicted pseudogap caused by the appearance of the spinon pairing order $\Delta^f$ decreases in the overdoped region and this, in turn, causes a decrease in holon pairing $\Delta^b$ and 
the coupling effect between the charge (holon-pair) and spin (spinon pair) degrees of freedom in the overdoped region thus a decrease in both $T_c$ and $\frac{n_s}{m^*}$.
This coupling between $\Delta^f$ and $\Delta^b$ is exhibited in the last term of Eq.(\ref{eq:su2_holon_action}).
Such coupling appears in exactly identical manner in both the U(1) and SU(2) theory.
Thus the boomerang behavior of the SU(2) theory is qualitatively the same as that of the U(1) theory.

In summary, we have found the following qualitatively salient features for the doping and temperature dependences of the superfluid weight.
We demonstrated the nearly doping independence in the slope of superfluid weight vs. temperature in the underdoped region, thus satisfying the experimentally observed relation $\frac{n_s}{m^*}(x,T) = \frac{n_s}{m^*}(x,0) - \alpha T$ with $\alpha$, a constant.
On the other hand, in the overdoped region the doping dependence in the slope of the superfluid weight with temperature is correctly predicted in agreement with observation.
From the present study we found a tendency of the boomerang behavior in the locus of $T_c$ with the variation of hole doping and superfluid weight in the plane of $\frac{n_s}{m^*}(x,T \rightarrow 0)$ vs. $T_c$.
This is in qualitative agreement with the $\mu$-SR experiments.
The decreasing (boomerang) behavior of $T_c$  and $\frac{n_s}{m^*}$  in the overdoped region is attributed to the weakened coupling of holon pairs to the spin pairing order.

One(SHSS) of us acknowledges the generous supports of Korea Ministry of Education(HakJin Program 2002-2003) and the Institute of Basic Science Research (2002-2003) at Pohang University of Science and Technology.


\begin{minipage}[c]{9cm}
\begin{figure}
\vspace{0cm}
\epsfig{file=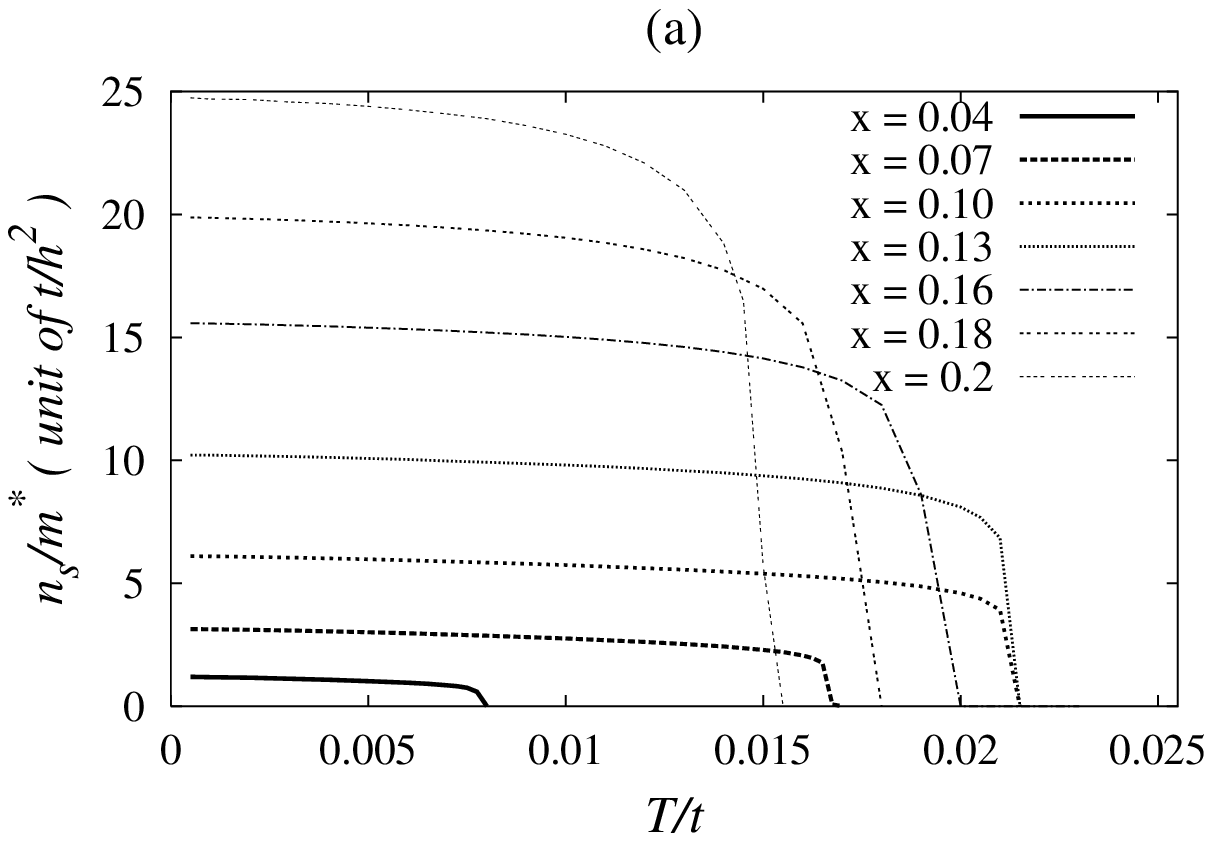,angle=0, height=6.9cm, width=6.9cm}
\end{figure}
 \end{minipage}
\\
\begin{minipage}[c]{9cm}
\begin{figure}
\vspace{0cm}
\epsfig{file=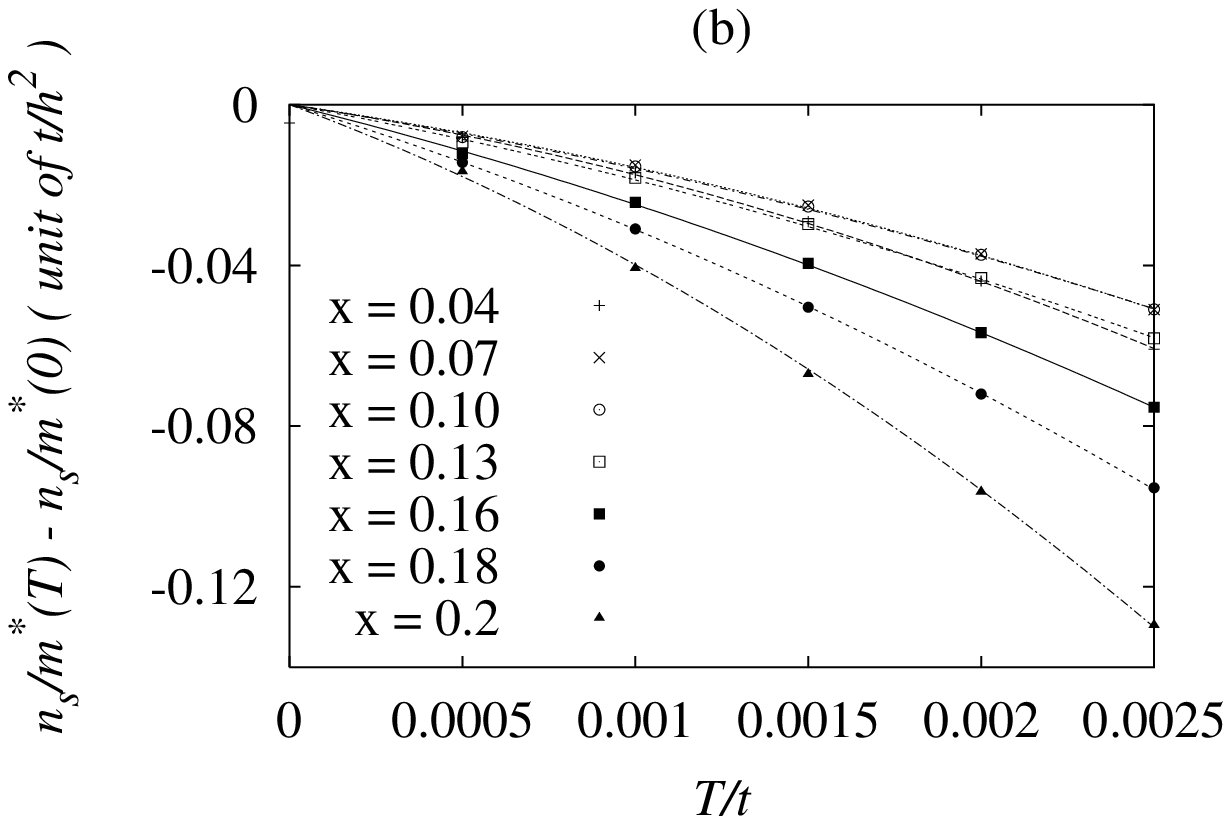,angle=0, height=6.9cm, width=6.9cm}
\end{figure}
 \end{minipage}
\\
\begin{minipage}[c]{9cm}
\begin{figure}
\vspace{0cm}
\epsfig{file=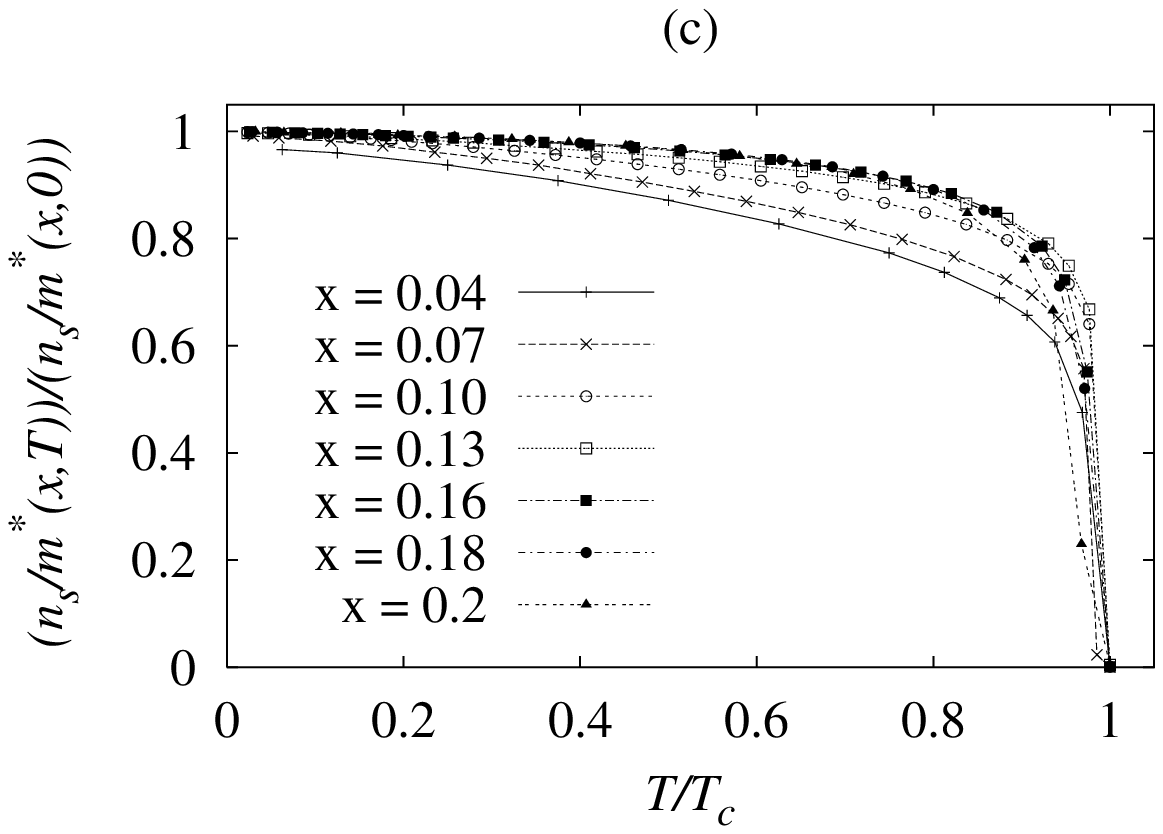,angle=0, height=6.9cm, width=6.9cm}
\label{fig:1}
\caption{
(a) Temperature dependence of the superfluid weight $\frac{n_s}{m^*}$ for underdoped($ x =0.04, 0.07, 0.1$), optimal doping($ x =0.13$) and overdoped($x = 0.16, 0.18, 0.2$) rates with $J/t=0.2$.  (b) The difference of superfluid weights between $T \neq 0K$ and $T=0K$, that is, $\frac{n_s}{m^*}(T) - \frac{n_s}{m^*}(T=0)$. Each line represents a fitting to the computed superfluid weight. (c) The normalized superfluid weight $\frac{n_s/m^*(x,T)}{n_s/m^*(x,0)}$ as a function of scaled temperature $T/T_c$ for same doping concentrations and $J$ value as (a).
}
\end{figure}
 \end{minipage}
\\
\\

\begin{minipage}[c]{9cm}
\begin{figure}
\vspace{0cm}
\epsfig{file=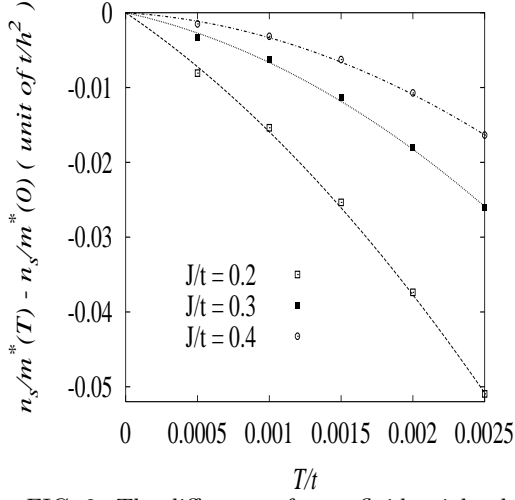,angle=0, height=6.9cm, width=6.9cm}
\label{fig:2}
\caption{
The difference of superfluid weights between $T \neq 0K$ and $T = 0K$, $\frac{n_s}{m^*}(T) - \frac{n_s}{m^*}(T=0)$ for a underdoped hole concentration $x = 0.1$ with $J/t=0.2, 0.3$ and $0.4$.  Each line represents a fitting to the computed superfluid weight.
}
\end{figure}
 \end{minipage}

\begin{minipage}[c]{9cm}
\begin{figure}
\vspace{0cm}
\epsfig{file=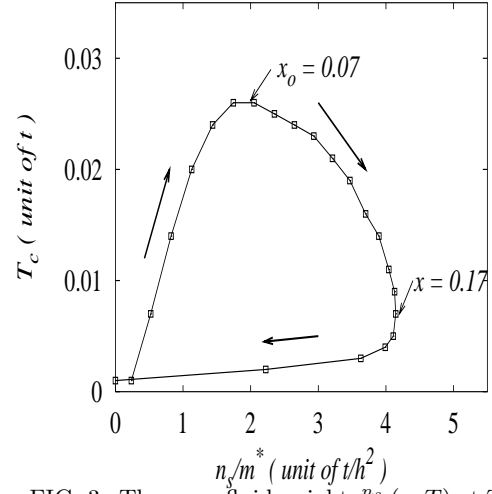,angle=0, height=6.9cm, width=6.9cm}
\label{fig:3}
\caption{
The superfluid weight $\frac{n_s}{m^*}(x,T)$ at $T=0.001t$ (equivalent to $T \sim 5K$ with the use of $t=0.44eV$[23]) vs. superconducting temperature $T_c$ with the choice of $J/t=0.2$ based on the U(1) theory.
The open box represents hole concentration starting from $x=0.01$ to $x=0.22$ and the arrow denotes the direction of increasing doping rate. 
The predicted optimal doping rate is $x_o = 0.07$ with the U(1) theory.
The SU(2) theory predicts qualitatively the same boomerang behavior.
}
\end{figure}
 \end{minipage}

\widetext
\end{document}